\documentclass[a4paper,11pt]{article}
\usepackage[dvipdfmx]{graphicx}%add [dvipdfmx]to repair graphics(sep.9,2014)
\usepackage{jheppub} % for details on the use of the package, please
\usepackage[T1]{fontenc} % if needed

\title{\boldmath 
Bosonization approach for 
``atomic collapse'' 
in graphene}

\author{Aya Kagimura,\note{Corresponding author.}}
\author{Tetsuya Onogi}

\affiliation{Department of Physics, Osaka University, Toyonaka, Osaka 560-0043, Japan}
\emailAdd{kagimura@het.phys.sci.osaka-u.ac.jp}
\emailAdd{onogi@phys.sci.osaka-u.ac.jp}

\abstract{We study quantum electrodynamics with 2+1 dimensional massless Dirac fermion around a Coulomb impurity. Around a large charge with atomic number Z > 137, the QED vacuum is expected to collapse due to the strong Coulombic force. While the relativistic quantum mechanics fails to make reliable predictions for the fate of the vacuum,  the heavy ion collision experiment also does not give clear understanding of this system. Recently, the ``atomic collapse'' resonances were observed on graphene where an artificial nuclei can be made. In this paper, we present our nonperturbative study of the vacuum structure of the quasiparticles in graphene with a charge impurity which contains multi-body effect using bosonization method. 
}

\begin{document} 
%\flushright{YITP-14-109, OU-HET-847}
\hspace{10cm} OU-HET-865
\maketitle
%\flushright{YITP-14-109, OU-HET-847}
%\flushright{YITP-14-109, OU-HET-865}
\flushbottom

%Start revising TO Jul 31,  end revising
\newpage 
\section{Introduction}
	The quantum electrodynamics (QED) is the most precisely experimentally tested theory in today's fundamental theories. In usual perturbation theory of QED, the expansion parameter is the fine structure constant $\alpha \sim \frac{1}{137}$. However, in strong external field, since the interaction is correspondingly strong, the perturbation theory breaks down. From the result of relativistic quantum mechanics, the vacuum around an atom with large atomic number $Z\gtrsim 137$ is expected to collapse. 
	
	In non relativistic quantum mechanics, in the region where the potential is larger than the energy, the wave function falls off exponentially. Namely, all the incoming particles are reflected; that is, the reflection rate is $R=1$ and transmision rate is $T=0$. On the other hand, in relativistic quantum mechanics, when height of the potential $V_0$ is larger than twice of particle mass $2m$, the reflection rate becomes larger than unity ($R>1$), which is called the Klein tunneling\cite{Klein:1929zz,Bjorken:1964}. This mechanism originates from the fact that the Dirac equation has both the positive and negative energy solutions as opposed to the Schr\"dingier equation. The same mechanism also prevents the electron to form bound states in a very strong attractive potential. 
%following peculiar result is known. The incoming particle that is positive energy solution of relativistic Shr\"odinger equation connect with negative energy solution of the equation in the potential, and go through the high potential barrier. Furthermore reflection rate becomes $R>1$. This phenomenon is called Klein's paradox\cite{Klein:1929zz,Bjorken:1964}. If the result is correct, the electron do not make bound states in very deep potential. 
In particular, an electron around a nuclei with a sufficiently large atomic number $Z$ does not form a bound state due to the strong Coulomb potential, and fall into the nuclei. This phenomenon is called the atomic collapse, and has been known theoretically for a long time. However, since the atom with $Z\gtrsim 137$ can be created for only a short time in heavy ion collision experiment, it is difficult to observe the phenomena experimentally at the quantitive level \cite{Schweppe:1983yv,Cowan:1985cn}. 
%	\begin{figure}[htbp]
%		\begin{minipage}{0.5\hsize}
%			\centering
%			\includegraphics[width=5cm]{nonrela.pdf}
%			\caption{}
%			\label{nonrela}
%		\end{minipage}
%		\begin{minipage}{0.5\hsize}
%			\centering
%			\includegraphics[width=5cm]{rela.pdf}
%			\caption{}
%			\label{rela}
%		\end{minipage}
%	\end{figure}
	
	The situation has changed since the discovery of the graphene in 2004\cite{Novoselov:2004}. The electric structure of the graphene at low energy is known to be the same as that of the massless Dirac fermion. In addition, the effective fine structure constant $\alpha $ is about $300$ times as large as that in the Quantum Electro Dynamics (QED). Due to this property, the essential point of the physics in the strongly coupled QED can be tested in the experiment using the graphene. Putting charged impurities on the graphene, one can realize a system similar to the large $Z$ atom system, which enable us to observe the ``atomic collapse''.
	
	The graphene is very thin, very light, very strong, and has very high electron conductivity and made from carbon atoms which is a ubiquitous element on earth. Therefore the graphene is expected to serve as ideal device in future. In this point of view, understanding the response of the electron to the charged impurity in graphene is a very important problem and is studied actively. This system is well studied in one body quantum mechanics as the system of the two dimensional massless electron in Coulomb potential \cite{Pereira:2007,Shytov:2007a,Nishida:2014}. It is predicted that when the charge of the impurity exceeds a critical value $Z_{cr}$, the wave function drastically changes. The massless fermion forms infinite number of quasibound states with negative energy, and the characteristic resonances appear in the local density of states (LDOS) of the electron\cite{Shytov:2007b}. Inspired by these theoretical studies, the scanning tunneling microscope (STM) experiment was carried out and a characteristic peak in LDOS was measured\cite{Wang:2013}.
	
	However, the above theoretical studies do not take into account the many body effect which involve electron-positron pair creation. In the graphene case, since the pair creation can occur with no cost of extra energy, the many body effect should not be neglected, which should be treated in the quantum field theory. Moreover, because of the large coupling perturbative approximation cannot be valid.
%the effect of the interaction between pair created electron and the impurity charge is very large. The effect can be contained in the quantum field theory. However, perturbative calculation is not available in strong coupled QED. Therefore, this problem should be calculated
Thus, this problem should be studied in some nonperturbative way. 
	
	We analyze the field theory of 2+1 dimensional Dirac massless fermion around an external charge using the bosonization technique. In two dimensional theory, the fermion theory is converted to the boson theory\cite{Coleman:1974bu,Mandelstam:1975hb}. It is known that a part of quantum effect of the fermion theory can be extracted from the classical boson theory. The bosonization method has been used to analyze the system with the fermion around monopole assuming that the classical boson theory captures the essential features of the quantum effect of the original fermion theory\cite{Callan:1982au,Harvey:1983tp}. The bosonization method is applied also to the atomic collapse problem in 3+1 dimensions\cite{Hirata:1986yt}. We apply this method to the atomic collapse problem in 2+1 dimensions.
	
	Following the studies in 3+1 dimensions mentioned above, first restricting the gauge and the fermion field to s-wave field, we reduce the theory to 1+1 dimensional fermion effective theory. Next, we map the two dimensional fermion theory to the two dimensional boson theory. Then we solve the classical equation of motion for the boson field. As a result, we find the vacuum structure including the charge screening of the impurity charge. %We also obtain induced electron density around the impurity. 
	
	This paper is organized as follows. In section 2, the result of foregoing analysis in one body theory for the Coulomb impurity problem on graphene is briefly reviewed. In section 3, we will explain the s-wave approximation and the bosonization formalism proposed in Ref. \cite{Hirata:1986yt}. In section 4, we will show the details about our study of vacuum solution and the result of our numerical analysis. Section 5 is devoted to summary and discussion. 
%May 14 revised by AK
% May 19 start revising  onogi --> May 19 end revising onogi
% May 27 start revising onogi  --> May 27 end revising onogi
% June 6 start revising kagimura  --> June 6 end revising kagimura
%%%%%%%%%%%%%%%%% BEGINING OF MACROS %%%%%%%%%%%%%%%%%%%%%%
\newcommand{\nn}{\nonumber}
\def\dfrac#1#2{\displaystyle\frac{#1}{#2}}
\newcommand{\ovl}[1]{\overline{#1}}
\newcommand{\wt}[1]{\widetilde{#1}}
\newcommand{\eq}[1]{Eq.~(\ref{#1})}
\newcommand{\eqn}[1]{(\ref{#1})}
\newcommand{\p}{\partial}
\newcommand{\pslash}{p\kern-1ex /}
\newcommand{\qslash}{q\kern-1ex /}
\newcommand{\lslash}{l\kern-1ex /}
\newcommand{\sslash}{s\kern-1ex /}
\newcommand{\kaslash}{k_a\kern-2ex /}
\newcommand{\kbslash}{k_b\kern-2ex /}
\newcommand{\Dslash}{{\cal D}\kern-1.5ex /}
\newcommand{\bpsi}{\overline{\psi}}
\newcommand{\bc}{\overline{c}}
\newcommand{\tr}{{\rm tr}}
\newcommand{\vev}[1]{\langle #1 \rangle}
\newcommand{\VEV}[1]{\left\langle{\rm T} #1\right\rangle}
\newcommand{\beqa}{\begin{eqnarray}}
\newcommand{\eeqa}{\end{eqnarray}}
%%%%%%%%%%%%%% END OF MACROS %%%%%%%%%%%%%%%%%%

\section{Review on Coulomb impurity on graphene}
\label{sec:model}
In this section, we review the Coulomb impurity problem on graphene. The electronic properties of the graphene are described by the tight-binding model where interactions between different orbits are neglected. And it is assumed that the electron can hop to only the nearest neighbor site. In momentum space, the energy of electron becomes zero at two points ($K$ and $K'$). The low energy effective theory is obtained by expanding the equation which the electron obeys around these points. It is known that the effective Hamiltonian takes the same form as that of massless Dirac fermion. That is, the fermionic low energy excitation obeys the Dirac equation
\begin{align}
	-v_F\left( 
	\begin{array}{cc}
		0 & \hat{p_x}-i\hat{p_y} \\
		\hat{p_x}+i\hat{p_y} & 0
	\end{array}
	\right) \psi =\varepsilon \psi ,	\label{Dirac equation}
\end{align}
and has the linear dispersion relation 
\begin{align}
	\varepsilon =\pm v_F\sqrt{p_x^2+p_y^2},	\label{linear dispersion relation}
\end{align}
at low energy. The parameter $v_F$ in the above equation is the Fermi velocity which is roughly estimated as $v_F\sim \frac{c}{300}$. Since $v_F$ plays the similar role as the speed of light $c$ in quantum electrodynamics, the effective fine structure constant for the fermionic excitations on graphene is $\alpha _{\rm eff}\sim \frac{300}{137}$. This means that the massless Dirac fermion on graphene is strongly coupled.

The behavior of electron in a hydrogen like atom is studied in relativistic quantum mechanics. It is known that the bound state of electron and a point charge $Ze$ cannot exist when $Z\alpha \ge 1$. For such a strongly coupled system, it is expected that the strong electric field makes the vacuum unstable since the strong Coulomb potential causes particle-hole pair creations. Such a phenomenon is called the ``atomic collapse'' and has been discussed for a long time. In the experimental side, the atomic collapse has been tested in heavy-ion collision. However the instability of large atomic number nuclei makes it difficult to observe the phenomenon clearly.

In the graphene case, such a situation can be easily set up due to the large value of the effective coupling $\alpha \sim \frac{300}{137}$ of the Dirac fermion. Recently, Wang and his collaborators studied the graphene system with Coulomb impurities with STM and observed the resonance like the quasibound state\cite{Wang:2013}. They put Ca dimers as impurity, and measured the local density of states (LDOS) of electron around the impurity. They showed that the peak appears in energy dependence of LDOS. The peak point is below the Dirac point when 5 Ca dimers are put. According to them, this is the quasi-bound state expected in one body theory. The quasi-bound state spatially spread through about 10 nm around  the center of Ca dimers in this experiment. 

In view of this STM experiment, it is now very important to study the graphene system with Coulomb impurities theoretically. In one body theory, the solution of the Dirac equation with Coulomb potential by a charged impurity can be exactly obtained\cite{Pereira:2007,Shytov:2007a}. The behavior of the solution drastically changes when $Z\alpha >1/2$. Because the electrons in graphene are massless, they do not seem to make bound state even in small $Z\alpha $. However, by introducing graphene lattice cutoff, the quasi-stable bound state is predicted to appear in strong coupling case. 

In Ref. \cite{Shytov:2007b}, the existence of the quasi-bound state is semiclassically discussed. Here, we briefly review their discussion. The Hamiltonian for 2 dimensional massless fermion in Coulomb potential is 
\begin{align}
	H={\boldsymbol \sigma }\cdot {\bf p}-\frac{Z\alpha }{r}.	\label{2DDhamiltonianiC}
\end{align}
When we write the square of momentum in terms of the radial momentum $p_r$ and the angular momentum $j$
\begin{align}
	p^2=p_r^2+j^2/r^2,
\end{align}
the Hamiltonian (\ref{2DDhamiltonianiC}) leads to
\begin{align}
	p^2_r=\left( \varepsilon +\frac{Z\alpha +j}{r}\right) \left( \varepsilon +\frac{Z\alpha -j}{r}\right) ,
\end{align}
where $\varepsilon $ is energy eigenvalue. The classically forbidden region where $p_r^2<0$ corresponds to
\begin{align}
	r_1\equiv \frac{Z\alpha -j}{|\varepsilon |}<r<\frac{Z\alpha +j}{|\varepsilon |}\equiv r_2.
\end{align}
Notice that if $Z\alpha >j$, there exist classically allowed region inside; that is, $r<r_1$. Therefore in strongly coupled case, quasibound states can be found by imposing the Bohr-Sommerfeld quantization condition 
\begin{align}
	\int _{r_0}^{r_1}p_r dr=n\pi ,
\end{align}
where $r_0$ is the lattice cutoff. 

In one particle theory, the interesting feature mentioned above can be found and LDOS can be calculated. However, since the atomic collapse is a phenomenon which comes from pair creation effect, it should be analyzed in a way which contain  nonperturbative multi body effects. In the following section, we will show the 2+1 dimensional massless fermion version of the bosonization formulation proposed in Ref. \cite{Hirata:1986yt}. 
%May 14 revised by AK
%May 15 revived by AK
%May 18 revived by AK
%May 22 start revising AK, end revising AK
%June 4 start revising TO 
%July 6 start revising AK,end revising AK
\newcommand{\Slash}[1]{{\ooalign{\hfil/\hfil\crcr$#1$}}} %Def of Feynman Slash
%%%%%%%%%%%%%%%%%%%%%%%%%%%%%%%%%%%%%%%%%%%%%%%%%%%%%%%%
\section{Approximation and Formalism}
\label{sec:formalism}
In this section, we study the vacuum structure of the massless Dirac fermion system in 2+1 dimensions around a Coulomb impurity. In order to analyze the system nonperturbatively, we employ the method proposed in Ref. \cite{Hirata:1986yt} for the atomic collapse QED in 3+1 dimensions. We firs restrict the theory with s-wave electromagnetic field and the lowest partial wave electron field. Under this approximation, the theory is reduced to 1+1 dimensional effective theory with time and radial degrees of freedom. We then bosonize the effective 1+1 dimensional fermion theory. Since it is known that the bosonized theory captures important part of the nonperturbative effect of the original fermion theory even at the classical level, we study the nonperturbative vacuum structure by constructing the classical solution of the bosonized theory. 
\subsection{1+1D Effective Theory}
Since the gauge field is in 3+1 dimensions, we start from the following gauge action
\begin{align}
	S_g=\int d^4 x \left[ -\frac{1}{4}F_{\mu \nu }F^{\mu \nu }-Ze\rho (x)A_0\right] ,	\label{starting gauge action}
\end{align}
where the charge density of impurity is spherically symmetric $\rho (x)=\rho (r,t)$, and normalized as $\int d^3x \rho (x)=1$. The s-wave electromagnetic field takes following form
\begin{align}
	A_0(x)=a_0(r,t),\ A_i(x)=\hat{r}_ia_1(r,t),	\label{s-waveEMF}
\end{align}
	where $\hat{r}_i=r_i/r$ is $i$th component of the unit vector in radial direction. In this approximation, the gauge action becomes
\begin{align}
	S_g=\int drdt\left[ 2\pi r^2(\partial _0a_1-\partial _ra_0)^2-4\pi Zer^2\rho (r,t)a_0\right] \label{Gaction}.
\end{align}	

When the graphene is on $z=0$ surface and the electron is trapped on this surface, the action for the electron coupled with the gauge field is 
\begin{align}
	S_f&=\int d^4x \left[ \overline{\psi } (i \Slash{\partial } +e\Slash{A})\psi \right] \delta (z)	\label{S3Df}\\
	&=\int d^4x \left[ \psi ^\dag \gamma ^0(i\gamma ^0\partial _0+i\gamma ^i\partial _i+e\gamma ^0A_0+e\gamma ^iA_i)\psi \right] \delta (z).\nonumber 
\end{align}
where $\psi $ is 2 component Weyl spinor. We take gamma matrices as 
\begin{align}
	\gamma ^0=\sigma _3,\ \gamma ^1 =i\sigma _2,\ \gamma ^2=-i\sigma _1. \label{gammaM}
\end{align}
Because we are considering $z=0$ surface and using the s-wave approximation (\ref{s-waveEMF}), $\gamma ^3$ disappears from (\ref{S3Df}). From now on, $i$ runs from $1$ to $2$. The fermion action becomes 
\begin{align}
	S_f=\int d^2xdt\psi ^\dag \left[ (i\partial _0+ea_0)+\sigma ^i(i\partial _i+e\hat{r}_ia_1)\right] \psi .
\end{align}

We expand the fermion field $\psi $ as
\begin{align}
	\psi =\frac{1}{\sqrt{r}}\sum_{m,\sigma }v_{m,\sigma }(r,t)\Psi _{m,\sigma }(\varphi ),
\end{align}
where $\sigma =\pm1$ and $m$ is half integer, and
\begin{align}	
	\Psi _{m,\sigma }=\frac{1}{\sqrt{4\pi }}\left( 
	\begin{array}{c}
		e^{i(m-1/2)\varphi }\\
		\sigma e^{i(m+1/2)\varphi }
	\end{array}
	\right) . 	\label{definition of Psi}
\end{align}
is normalized as 
\begin{align}
	\int d\varphi \Psi_{m'\sigma '}^\dag \Psi_{m,\sigma }=\delta _{m,m'}\delta _{\sigma ,\sigma '}.
\end{align}
Using the relation 
\begin{align}
		\sigma _i\hat{r}_i\Psi _{m,\sigma }=\sigma \Psi _{m,\sigma },
\end{align}
we get 
\begin{align}
	\sigma _i\partial _i\psi =\frac{1}{\sqrt{r}}\sum _{m,\sigma }\sigma \left( \partial _rv_{m,\sigma }(r,t)\Psi _{m,\sigma }(\varphi )+\frac{m}{r}\Psi _{m,-\sigma }(\varphi )\right) .
\end{align}
Therefore the action for fermion becomes
\begin{align}
	S_f=\int drdt \sum_{m,\sigma }\left[ v_{m,\sigma }^*\left\{ i\partial _0+ea_0+\sigma (i\partial _r+ea_1)\right\} v_{m,\sigma }-i\sigma v_{m,\sigma }^*\frac{m}{r}v_{m,-\sigma }\right] .
\end{align}
	
We restrict ourself to consider only the lowest ($j=1/2$) partial wave, and define 1+1 dimensional fermion 
\begin{align}
	u_m&:=\left( \frac{1+i}{2}+\frac{1-i}{2}\sigma _3\right) \left( 
	\begin{array}{c}
		v_{m,+}\\
		{\rm sign} (m) v_{m,-}
	\end{array}
	\right) \nonumber \\
	&=\left( 
	\begin{array}{c}
		v_{m,+}\\
		{\rm sign}(m)i v_{m,-}
	\end{array}
	\right) , 
\end{align}
where $m=\pm 1/2$. From now on, we take 
\begin{align}
	\gamma ^0=\sigma _2,\ \gamma ^1=i\sigma _1,\ \gamma _5 =\gamma ^0\gamma ^1=\sigma _3
\end{align}
as 2 dimensional gamma matrices. Then we can rewrite 2 dimensional fermion action as
\begin{align}
	S_f=\int drdt\sum_{m=\pm 1/2}\left[ \overline{u}_m\left\{ \gamma ^0(i\partial _0+ea_0)+\gamma ^1(i\partial _r+ea_1)\right\} u_m+i\frac{1}{2r}\overline{u}_m\gamma ^5u_m\right] 	\label{2DFactionIG}.
\end{align}
The last term represents centrifugal force. Unlike that of Ref. \cite{Hirata:1986yt},we have a different coefficient of centrifugal force term and no mass term. 
	
We have to set the boundary condition for fermion field $u_{m}$ at $r=0$ by requiring no singularity at $r=0$. From (\ref{definition of Psi}), 
\begin{align}
	\Psi _{1/2,+}-\Psi _{1/2,-},\ \Psi _{-1/2,+}+\Psi _{-1/2,-}
\end{align}
has $\varphi $ dependence. If the coefficients of these are finite value at $r=0$, the singularity arises. So, we set the boundary condition 
\begin{align}
	v_{m,+}(0,t)-{\rm sign}(m)v_{m,-}(0,t)=0. 
\end{align}
Written in 2D fermion $u_m$, 
\begin{align}
	(1-\gamma ^0)u_m(0,t)=0. 	\label{2dfboundary}
\end{align}
On the other hand, since 
\begin{align}
	\Psi _{1/2,+}+\Psi _{1/2,-},\ \Psi _{-1/2,+}-\Psi _{-1/2,-}
\end{align}
don't have $\varphi $ dependence, the coefficient of these can be finite at $r=0$. Therefore we can also use the same boundary condition as Ref. \cite{Hirata:1986yt}.
	
By the way, in one body theory, the boundary condition is set not at $r=0$, but at $r=r_0$\cite{Pereira:2007, Shytov:2007a}, which is lattice cut off size of graphene. And the cut off plays very important role to discuss the drastic change of wave function and quasi-bound state in strong coupling region. In our case, however, even if we set the boundary condition at $r=r_0$, we get the same result for $r=0$. Therefore, here we set the boundary condition at $r=0$. 
%%%%%%%%%%%%%%%%%%%%%%%%%%
\subsection{Bosonization}
We apply bosonization to this theory. Regarding interaction term as perturbation, we bosonize free fermion field to free boson field, 
\begin{align}
	u_m(r,t)=\left( \frac{\mu }{2\pi }\right) ^{1/2}\left( 
	\begin{array}{c}
		-iN_\mu \exp [i\sqrt{\pi }(\phi _m(r,t)+\tilde{\phi }_m(r,t))]\\
		N_\mu \exp [i\sqrt{\pi }(-\phi _m(r,t)+\tilde{\phi }_m(r,t))]
	\end{array}
	\right) ,
\end{align}
where 
\begin{align}
	\tilde{\phi }(x)=\lim_{\epsilon \to 0}\int_r^\infty dse^{-\epsilon s}\dot{\phi }(s,t),	\label{phitilde}
\end{align}
and $N_\mu $ represents normal ordering at IR mass scale $\mu $. From now on, we use the overdot and prime for time and spatial derivative, respectively. Because the action and the boundary condition are almost the same as Ref. \cite{Hirata:1986yt}, we can bosonize this theory following the same calculation. 

In this case, we should impose the boundary condition on the boson field. The boundary condition (\ref{2dfboundary}) is rewritten in boson field as
\begin{align}
	\phi _m(0,t)=0.	\label{BboundaryC}
\end{align}
Free boson field can be expanded in plane wave as
\begin{align}
	\phi (x,t)=\int_{k>0} \frac{dk}{2\pi }\left[ \bar{a}^\dag (k)e^{ik(x+t)}+a(k)e^{ik(x-t)}+\bar{a}(k)e^{-ik(x+t)}+a^\dag (k)e^{-ik(x-t)}\right], 
\end{align}
where $a,\ \bar{a},\ a^\dag ,\ \bar{a}^\dag $ are creation-annihilation operators satisfying appropriate commutation relations. While $a(k),\bar{a}(k)$ are independent operators without the boundary condition, with the boundary condition (\ref{BboundaryC})
\begin{align}
	0&=\phi (0,t)\nonumber \\
	&=\int_{k>0} \frac{dk}{2\pi }\left[ (\bar{a}^\dag (k)+a^\dag (k))e^{ikt}+(a(k)+\bar{a}(k))e^{-ikt}\right] ,
\end{align}
these are dependent on each other
\begin{align}
	\bar{a}(k)=-a(k).
\end{align}
Then the boson field $\phi $ and $\tilde{\phi }$ can be written as
\begin{align}
	\phi (r,t)=\int _{k>0} \frac{dk}{2\pi }\left[ a(k)(e^{ikr}-e^{-ikr})e^{-ikt}+a^\dag (k)(e^{-ikr}-e^{ikr})e^{ikt}\right] ,
\end{align}
and 
\begin{align}
	\tilde{\phi }(r,t)=\int _{k>0}\frac{dk}{2\pi }\left[ \left( e^{ikr}+e^{-ikr}\right)a(k)e^{-ikt}+\left( e^{ikr}+e^{-ikr}\right)a^\dag (k)e^{ikt} \right] .
\end{align}
We split these into 
\begin{align}
	\phi ^{(+)}(r,t)=\int _{k>0} \frac{dk}{2\pi }a(k)\left( e^{ikr}-e^{-ikr}\right) e^{-ikt}, 
\end{align}
\begin{align}
	\phi ^{(-)}(r,t)=\int _{k>0} \frac{dk}{2\pi }a^\dag (k)\left( e^{-ikr}-e^{ikr}\right) e^{ikt}, 
\end{align}
\begin{align}
	\tilde{\phi }^{(+)}(r,t)=\int_{k>0}\frac{dk}{2\pi }a(k)\left( e^{ikr}+e^{-ikr}\right) e^{-ikt}, 
\end{align}
\begin{align}
	\tilde{\phi }^{(-)}(r,t)=\int _{k>0}\frac{dk}{2\pi }a^\dag (k)\left( e^{ikr}+e^{-ikr}\right)e^{ikt}.
\end{align}
From the commutation relation $[a(k),a^\dag (k)]=2\pi \frac{1}{2k}\delta (k-k')$, we get the relation
\begin{align}
	&[\tilde{\phi }^{(+)}(r,t)+\eta \phi ^{(+)}(r,t),\tilde{\phi }^{(-)}(r',t')+\eta '\phi ^{(-)}(r',t')]\nonumber \\
	&\qquad =-\frac{1}{4\pi }[(1-\eta )(1-\eta ')A_++(1+\eta )(1+\eta ')A_-+(1-\eta )(1+\eta ')B_++(1+\eta )(1-\eta ')B_-],	\label{BcommutationR}
\end{align}
where 
\begin{align}
	A_\pm (r,t;r',t')&\equiv -\int_{k>0}dk\frac{1}{k}e^{ik(\mp (r-r')-(t-t'))}\nonumber \\
	&=\lim_{\epsilon \to 0}\ln \left( i\mu [ t-t'\pm (r-r')-i\epsilon \right] )	\label{A},
\end{align}
\begin{align}
	B_\pm (r,t;r',t')&\equiv -\int_{k>0}dk\frac{1}{k}e^{ik(\mp (r+r')-(t-t'))}\nonumber \\
	&=\lim_{\epsilon \to 0}\ln \left( i\mu [ t-t'\pm (r+r')-i\epsilon \right] )	\label{B},
\end{align}
are renormalized at IR mass scale $\mu $. $B_{+,-}$ arise from the boundary condition.

Using the commutation relation (\ref{BcommutationR}), we rewrite the interaction terms in fermion theory in terms of boson field. After some point splitting procedure, we get
\begin{align}
	\overline{u}\gamma ^\mu u=-\frac{1}{\sqrt{\pi }}\epsilon ^{\mu \nu }\partial _\nu \phi 	\label{interactionT},
\end{align}
\begin{align}
	\overline{u}\gamma _5u=-\frac{i}{2\pi r}N_\mu \cos (2\sqrt{\pi }\phi )	\label{centrifugalT},
\end{align}
where $\epsilon $ is anti-symmetric symbol with $\epsilon ^{10}=1$. 

We rewrite the fermion action (\ref{2DFactionIG}) in terms of the above boson operators. In $a_1=0$ gauge,
\begin{align}
	S_f=\int drdt\sum_{m=\pm 1/2}\left[ \frac{1}{2}\partial ^\mu \phi _m\partial _\mu \phi _m-\frac{e}{\sqrt{\pi }}a_0'\phi _m+\frac{1}{4\pi r^2}\cos (2\sqrt{\pi }\phi _m)\right] ,
\end{align}
where the second term is integrated by parts. Therefore we get the total action 
\begin{align}
	S=\int drdt\left[ 2\pi r^2a_0'^2+e\Phi (r,t)a_0'+\sum_{m=\pm 1/2}\left( \frac{1}{2}\partial ^\mu \phi _m\partial _\mu \phi _m-\frac{e}{\sqrt{\pi }}a_0'\phi _m+\frac{1}{4\pi r^2}\cos (2\sqrt{\pi }\phi _m)\right) \right] , 	\label{BTaction}
\end{align}
where the $\Phi (r,t)$ is defined by 
\begin{align}
	\Phi '(r,t)=4\pi Zr^2\rho (r,t).
\end{align}

From the action (\ref{BTaction}), we notice that $a_0$ has no dynamical degrees of freedom. Using the equation of motion for $a'_0$
\begin{align*}
	4\pi r^2a_0'+e\Phi (r,t)-\sum_{m=\pm 1/2}\frac{e}{\sqrt{\pi }}\phi _m=0,
\end{align*}
we can eliminate $a_0$. Therefore the Hamiltonian becomes
\begin{align}
	H=\int dr\left[ \sum_{m=\pm 1/2}\left\{ \frac{1}{2}\left( \pi _m ^2+\phi _m'^2\right)-\frac{1}{4\pi r^2}\cos (2\sqrt{\pi }\phi _m)\right\} +\frac{e^2}{8\pi r^2}\left( \Phi (r,t)-\frac{1}{\sqrt{\pi }}\sum_m \phi _m\right) ^2\right] .
\end{align}
Adding the c-number to the Hamiltonian, 
\begin{align}
	H&=\int dr\left[ \sum_{m=\pm 1/2}\left\{ \frac{1}{2}\left( \pi _m ^2+\phi _m'^2\right)+\frac{1}{4\pi r^2}\left( 1-\cos (2\sqrt{\pi }\phi _m)\right) \right\} \right. \nonumber \\
	&\qquad \left. +\frac{e^2}{8\pi r^2}\left\{ \left( \Phi (r,t)-\frac{1}{\sqrt{\pi }}\sum_m \phi _m\right) ^2-\Phi (r,t)^2\right\} \right] \label{hamiltonian},
\end{align}
we shift the energy so that the energy becomes zero when $\phi _m=0$ which is vacuum configuration with $Z=0$. In the next section, we numerically calculate the solution which minimize this Hamiltonian. 
%\input{4.Analysis.tex}
%May 14 revised by AK
%May 21 Start revising Kagimura, end revising AK
%May 22 Start revising Kagimura, end revising AK
%May 31 Start revising Kagimura, end revising AK
%June 3 Start revising Kagimura, end revising AK
%June 24 Start revising TO end revising TO
%July 24 Start revising TO end revising TO
%July 25 Start revising AK end revising AK
%July 27 Start revising AK end revising AK
%%%%%%%%%%%%%%%%% BEGINING OF MACROS %%%%%%%%%%%%%%%%%%%%%%
%%%%%%%%%%%%%% END OF MACROS %%%%%%%%%%%%%%%%%%

 \section{Study of Vacuum Solution}
 \label{sec:Study of Vacuum Solution}
In this section we find the classical solution which minimizes the bosonized Hamiltonian in the previous section. For this purpose, we have to solve the Euler-Lagrange equations for boson fields under the appropriate boundary conditions. The boundary condition at $r=0$ is determined by eq.(\ref{BboundaryC}). The solution is characterized by the boundary condition at $r=\infty $. 

Eq.(\ref{interactionT}) indicates that the density of electron $\rho _e(r)$ can be written in terms of the boson field as
\begin{align}
	\rho _e(r)&\equiv \psi ^\dag \psi =\sum_m \frac{1}{\sqrt{\pi }}\partial _r\phi _m(r).
\end{align}
Therefore, we get the spatial distribution of induced electron density corresponding to the solution. Total induced charge which screens the impurity charge is given by 
\begin{align}
	Q_{\mathrm{EM}}&\equiv -e\int dr \rho _e(r)=-\frac{e}{\sqrt{\pi }}\sum_m\phi _m(\infty ).
\end{align}
In order to study the vacuum structure, we consider only static solution $\pi _m=0$. We rewrite the Hamiltonian in terms of the new variable 
\begin{align}
	\phi _\pm =\frac{1}{\sqrt 2}(\phi _{+1/2}\pm \phi _{-1/2}).
\end{align}
Using the formula
\begin{align}
	\cos [\sqrt{2\pi }(\phi _++\phi _-)]+\cos [\sqrt{2 \pi }(\phi _+-\phi _-)]=2\cos (\sqrt{2\pi }\phi _+)\cos (\sqrt{2\pi }\phi _-),
\end{align}
the Hamiltonian becomes
\begin{align}
	H&=\int _0^\infty dr \left[ \frac{1}{2}(\phi '^2_++\phi _-'^2)+\frac{1}{2\pi r^2}\{ 1-\cos (\sqrt{2\pi }\phi _+)\cos (\sqrt{2\pi }\phi _-)\} \right. \nonumber \\
	&\qquad \left. +\frac{\alpha }{\pi r^2}\left\{ \left( \phi _+-\sqrt{\frac{\pi }{2}}\Phi (r,t)\right) ^2-\frac{\pi }{2}\Phi (r,t)^2\right\} \right] , 
\end{align}
where $\alpha =\frac{e^2}{4\pi }$. The Euler-Lagrange equations for $\phi _+, \phi _-$ are given by 
\begin{align}
	\phi _+''-\frac{1}{\sqrt{2\pi }r^2}\sin (\sqrt{2\pi }\phi _+)\cos (\sqrt{2\pi }\phi _-)-\frac{2\alpha }{\pi r^2}\left( \phi _+-\sqrt{\frac{\pi }{2}}\Phi (r)\right) =0,	\label{eq:phi+}
\end{align}
\begin{align}
	\phi _-''-\frac{1}{\sqrt{2\pi }r^2}\cos (\sqrt{2\pi }\phi _+)\sin (\sqrt{2\pi }\phi _-)=0,	\label{eq:Eqphi-}
\end{align}
respectively. Since it satisfies Eq.(\ref{eq:Eqphi-}) we can take the symmetric ansatz $\phi _-=0$. Then Eq.(\ref{eq:phi+}) reduces to
\begin{align}
	\phi _+''-\frac{1}{\sqrt{2\pi }r^2}\sin (\sqrt{2\pi }\phi _+)-\frac{2\alpha }{\pi r^2}\left( \phi _+-\sqrt{\frac{\pi }{2}}\Phi (r)\right) =0.	\label{ELEPhi+}
\end{align}
We assume that the impurity charge is spherically spread over radius $R$:
\begin{align}
	\rho (r)=\frac{3}{4\pi R^3}\theta (R-r).
\end{align}
The corresponding $\Phi (r)$ is 
\begin{align}
	\Phi (r)=\left\{ 
	\begin{array}{cc}
		Z\left( \frac{r}{R}\right) ^3&\quad (r<R)\\
		Z&\quad (r>R).
	\end{array}
	\right. 
\end{align}
%
%How is the boundary condition at $r=\infty $ determine? 
Since Eq.(\ref{ELEPhi+}) is a second order differential equation, in addition to the boundary condition at the origin we need to impose another boundary condition at $r=\infty $. For finiteness of total energy, the boson field should asymptotically be constant ($\phi_+\rightarrow \phi _*$) at large $r$. Substituting $\phi _+=\phi _*$ into the Euler-Lagrange equation at large $r$, we find that $\phi _*$ should be the solution of the following equation:
\begin{align}
	\sin (\sqrt{2\pi }\phi _*)=-2\alpha \left( \sqrt{\frac{2}{\pi }}\phi _*-Z\right) .	\label{ELEinfinity}
\end{align}
Notice that the asymptotic value $\sqrt{\frac{2}{ \pi }}\phi _*$ can take non-integer value. Charge screening with non-integer charge may seem counter intuitive if one tries to interpret the phenomena as particle hole pair creation. One should interpret such screening as the polarization effect.

In fact, it is known that the screening of non-integer charge actually occurs in massless Schwinger model \cite{Callan:1982au, Rubakov:1983vn,Iso:1988zi}.
%We can interpret the fractional screening not as electron positron pair creation, but as the polarization effect. 
In the following subsections, we show the detailed numerical analysis and its results. 
%%%%%%%%%%%%%%%%%%%%%%%%%%%%
\subsection{Numerical Analysis}

\subsubsection{Strategy}
Our numerical analysis is done in various parameters $\alpha , Z$, according to the following steps: 	
\begin{description}
	\item[(i)]\mbox{}Find the solution of Eq.(\ref{ELEinfinity}) and obtain the asymptotic form at large $r$.
	\item[(ii)]\mbox{}Solve the Euler-Lagrange equation (\ref{ELEPhi+}) with the boundary condition at large $r$ (\ref{asymptoticS}) with various $A$.
	\item[(iii)]\mbox{}Find $A$ with which the solution satisfies the boundary condition at $r=0$ (\ref{BboundaryC}). 
\end{description}

\subsubsection{Asymptotic form}
In order to numerically solve the Euler-Lagrange equation (\ref{ELEPhi+}), we should  find the asymptotic form at large $r$. 
To do so, we parameterize $\phi_+(r)$ by introducing a function $f$ which describes the deviation of $\phi _+(r)$ and $\phi_*$ at large $r$ as
\begin{align}
	\phi _+(r)=\phi _* - f(r).	\label{asymptoticf}
\end{align}
where $\phi _*$ is the solution of Eq.(\ref{ELEinfinity}). 
Substituting Eq.(\ref{asymptoticf}) into Eq.(\ref{ELEPhi+}),  and expanding it up to linear order in $f$, we obtain 
\begin{eqnarray}
f^{\prime\prime} -\frac{1}{r^2}\left( \cos(\sqrt{2\pi} \phi_*) + \frac{2\alpha}{\pi} \right) f+ O(f^2)= 0 
\label{eq:eq_f}
\end{eqnarray}
Assuming that the solution for $f$ can take the form 
\begin{align}
	f(r) \approx \frac{A}{r^\lambda },	\label{asymptotic_sol_f}
\end{align}
at large $r$ with $A$ being a constant and substituting it into Eq.(\ref{eq:eq_f}), we find that the power $\lambda$ satisfies the following equation:
\begin{eqnarray}
\lambda^2 + \lambda -\left(\cos(\sqrt{2\pi} \phi_*) + \frac{2\alpha}{\pi} \right) = 0 .
\label{eq:quad_lambda}
\end{eqnarray}

From Eq.(\ref{eq:quad_lambda}), $\lambda $ should be
\begin{align}
	\lambda =\frac{1}{2}\left[ -1+\sqrt{1+4\cos (\sqrt{2\pi }\phi _*)+\frac{8\alpha }{\pi }}\right] .	\label{eq:lambda}
\end{align}
\begin{figure}[htbp]
	\centering
	\includegraphics[width=8cm]{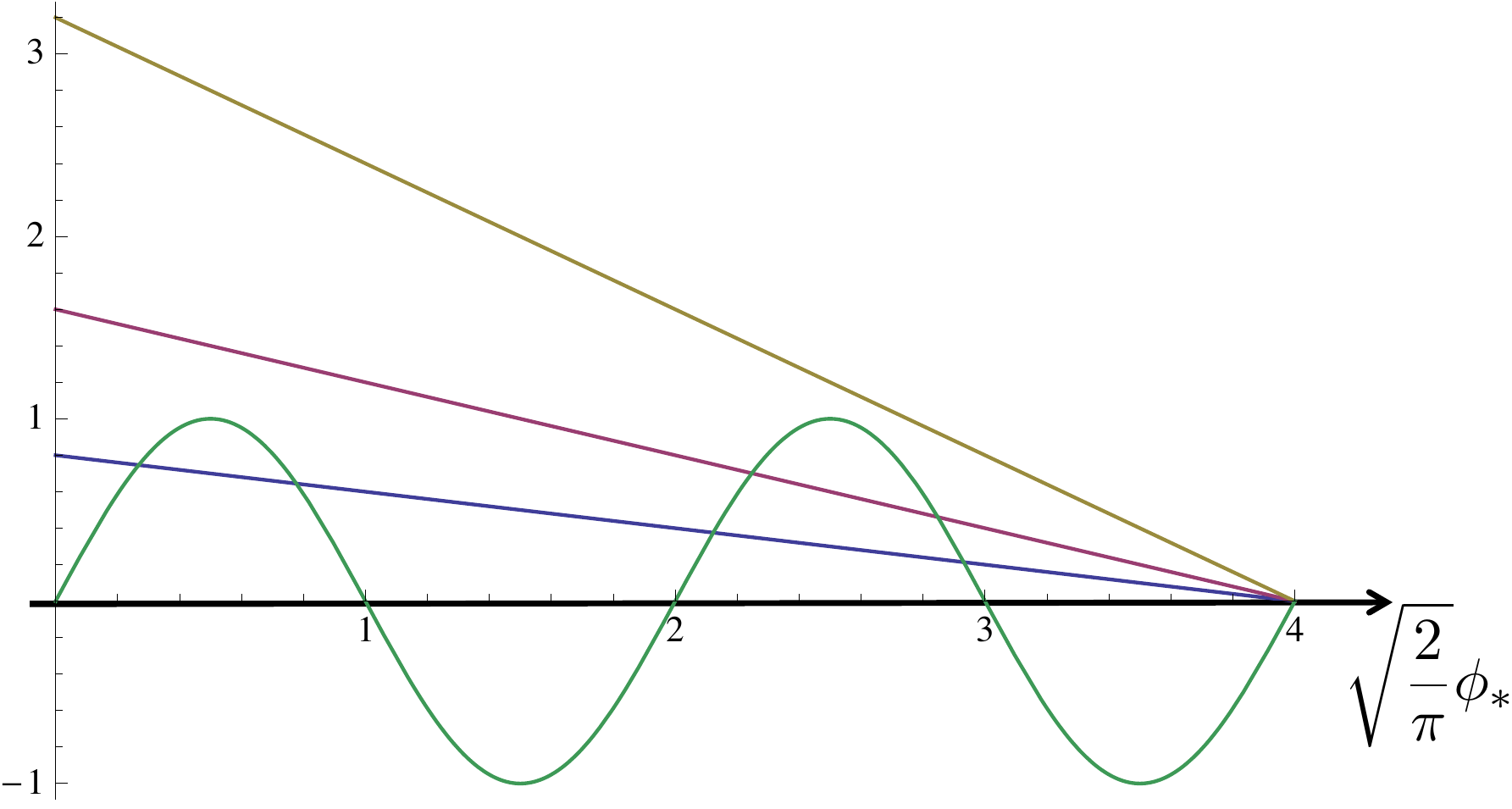}
		\caption{The green line is l.h.s. of eq.(\ref{ELEinfinity}), and blue, red, yellow lines are r.h.s. of eq.(\ref{ELEinfinity}) with $\alpha =0.1$, $\alpha =0.2$, $\alpha =0.4$, respectively.}
		\label{sin_linear}
\end{figure}
We show some examples of $Z=4$ case. In this case, there are three screening patterns depending on the value of $\alpha $ as shown in Fig. \ref{sin_linear}. In $\alpha \lesssim 0.14$ case, there are five values of $\phi_*$. However, when $\phi _*$ is equal to the second or fourth smallest value, based on Eq.(\ref{eq:lambda}), $\lambda $ becomes imaginary. Only the solutions with the real positive values of $\lambda $ make sense. So, there are three possibilities. For $0.14\lesssim \alpha \lesssim 0.34$ case, there are three values of $\phi_*$. Similarly the second smallest value of $\phi _*$ is not a physical solution. So, there are two possibilities. And in $0.34\lesssim \alpha $ case, there is only value for $\phi_*$ which corresponds to the full screening solution. 

\begin{figure}[b]
	\centering
	\includegraphics[width=12cm]{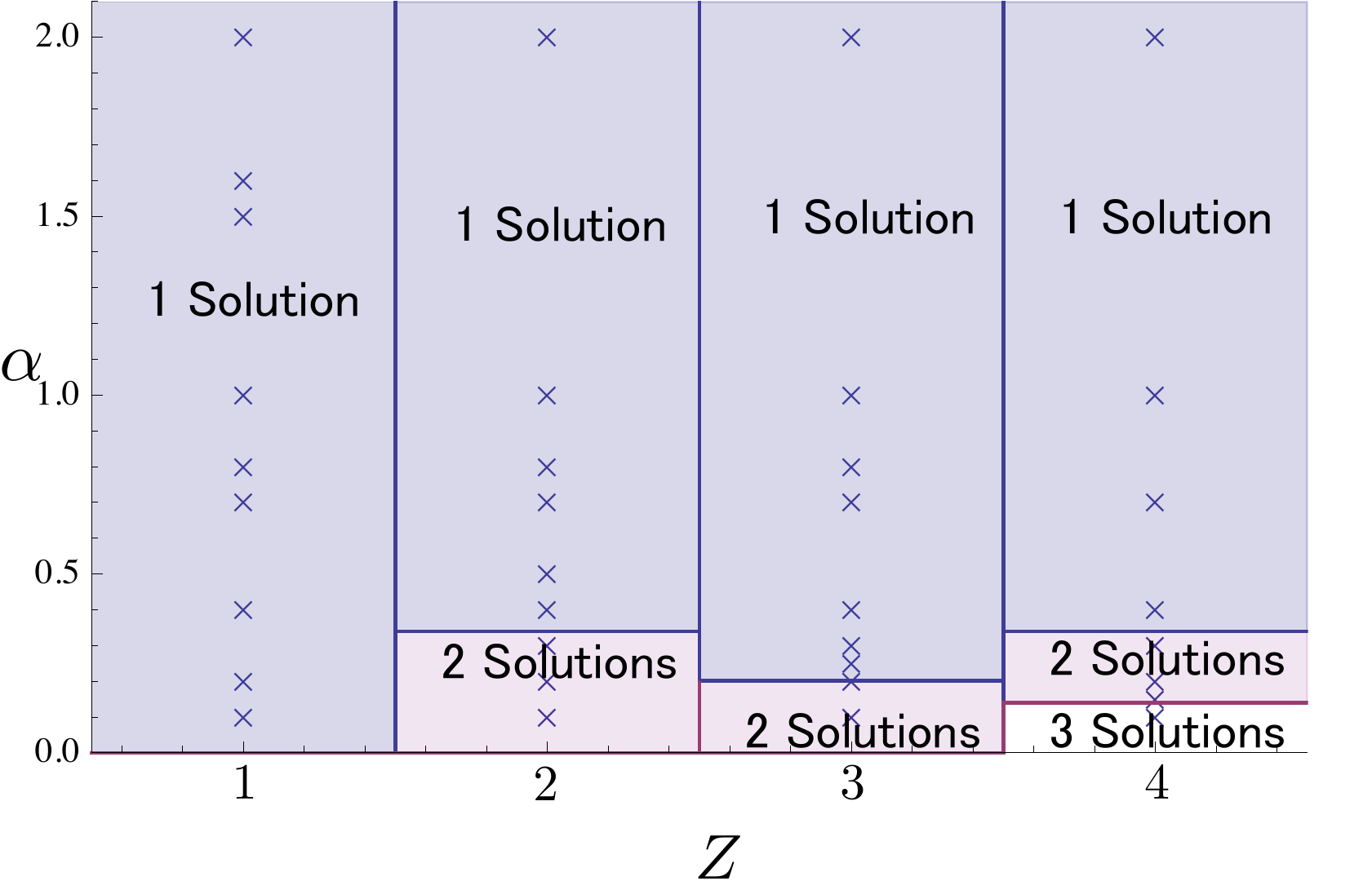}
	\caption{Number of possible  asymptotic solutions at large $r$ for each set of values of ($\alpha$, $Z$). Crosses are the parameter points where we solved Eq. (\ref{ELEPhi+})}
	\label{Number}
\end{figure}

In Fig. \ref{Number}, we show the number of possible asymptotic solutions at large $r$ for each set of values of ($\alpha$, $Z$).
%
%\begin{figure}[htbp]

\subsubsection{Example of the solution}
\begin{figure}[htbp]
	\begin{minipage}{0.5\hsize}
		\centering
		\includegraphics[width=7cm]{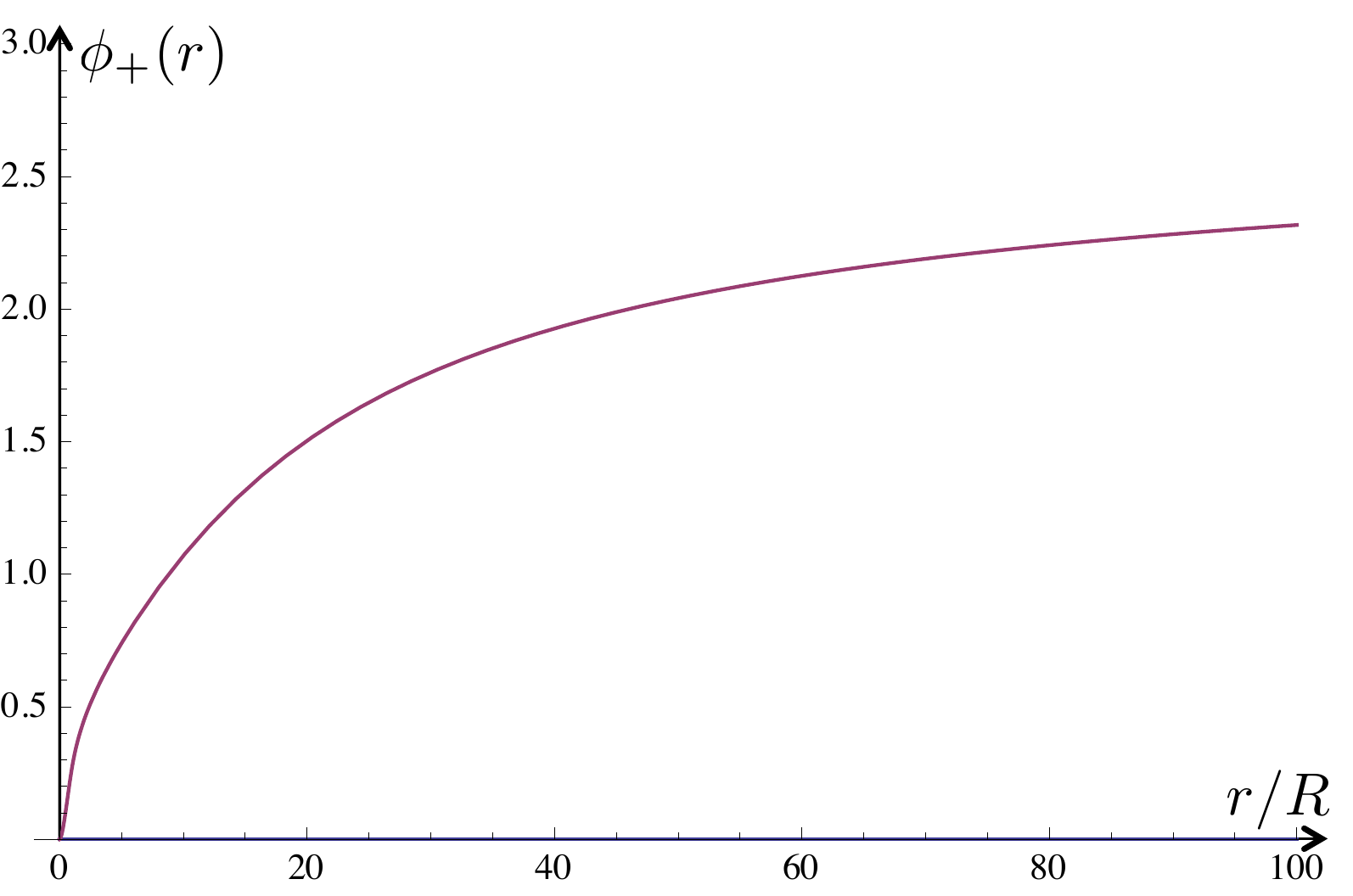}
		\caption{The solution for $\alpha =0.2, Z=4$.}
		\label{sol024}
	\end{minipage}
	\begin{minipage}{0.5\hsize}
		\centering
		\includegraphics[width=7cm]{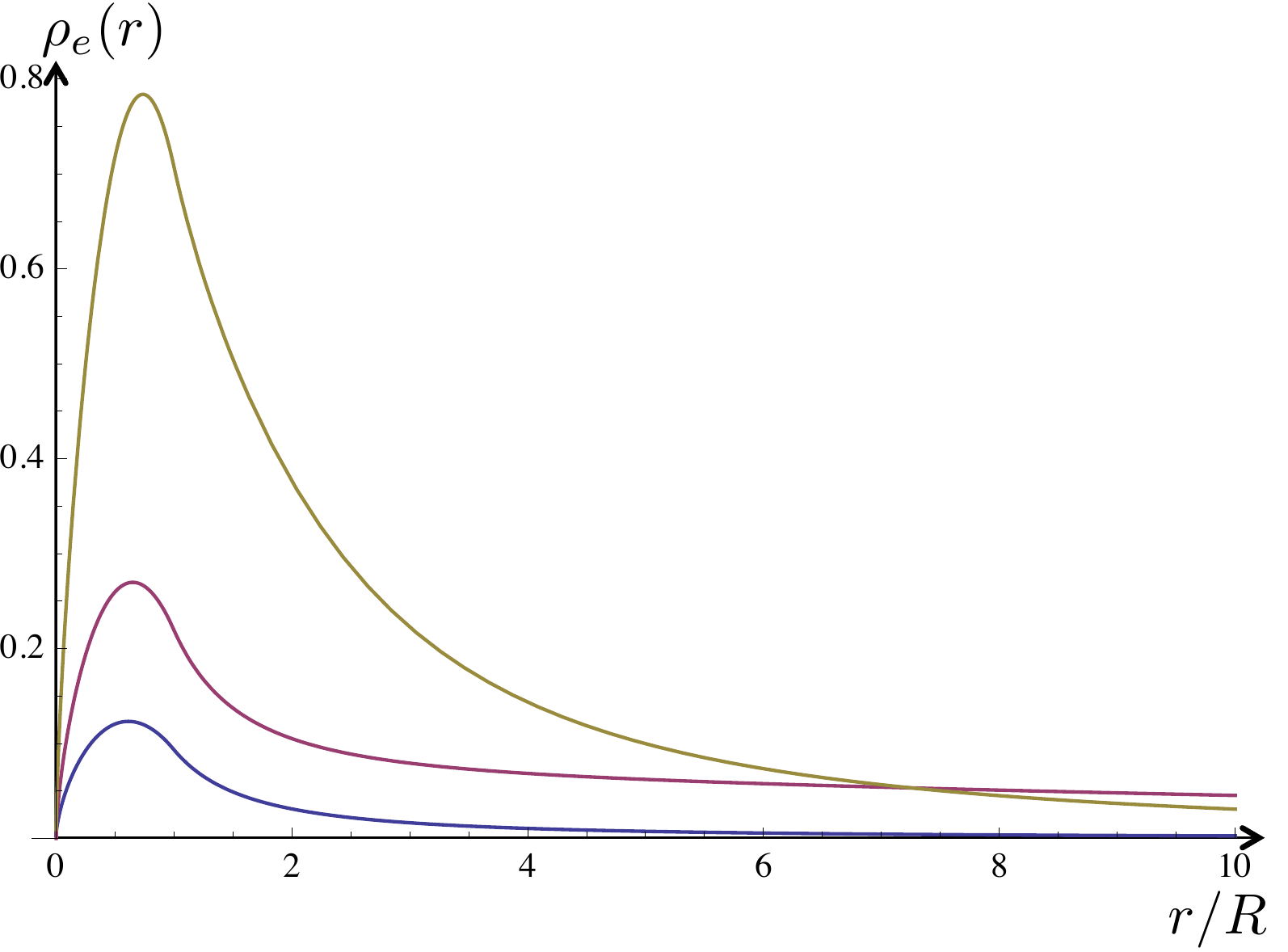}
		\caption{Induced electron density with $\alpha =0.1$ (blue line), $\alpha =0.2$ (red line), $\alpha =0.4$ (yellow line).}
		\label{density}
	\end{minipage}
\end{figure}

Starting from the asymptotic solutions and  solving the differential equation numerically, we can obtain the full solution.
Taking the following asymptotic from  
\begin{align}
	\phi _+(r) \approx \phi _*-\frac{A}{r^\lambda },	\label{asymptoticS}
\end{align}
at large $r$ and varying $A$, we can search for the physical solution which satisfies the boundary condition at $r=0$.
Practically, we solve Eq. (\ref{ELEPhi+}) from $r=0.001R$ to $r=100000R$, setting the boundary condition at large $r=100000R$ with various values of $\alpha, Z$.
For illustration, we show the example for $Z=4$ and $\alpha=0.2$.
In this case, there are two asymptotic solutions, but only the solution which realizes the smallest value of $\phi_*$ can satisfy appropriate boundary condition. 
The full solutions from the other asymptotic forms do not satisfy the boundary condition at $r=0$ but end up have positive values no matter how we choose the value of $A$. 
We show the solution of $\phi_+(r)$  in Fig. \ref{sol024}. In the other case, the shapes of solutions are qualitatively similar to the solution in this case. The induced electron density is depicted in Fig. \ref{density}. We notice that most of the induced electrons fall into the impurity. 
%%%%%%%%%%%%%%%%%%%%%%%%%%%%
\subsection{Result}
\subsubsection{Phase structure}
\begin{figure}[htbp]
		\centering
		\includegraphics[width=16cm]{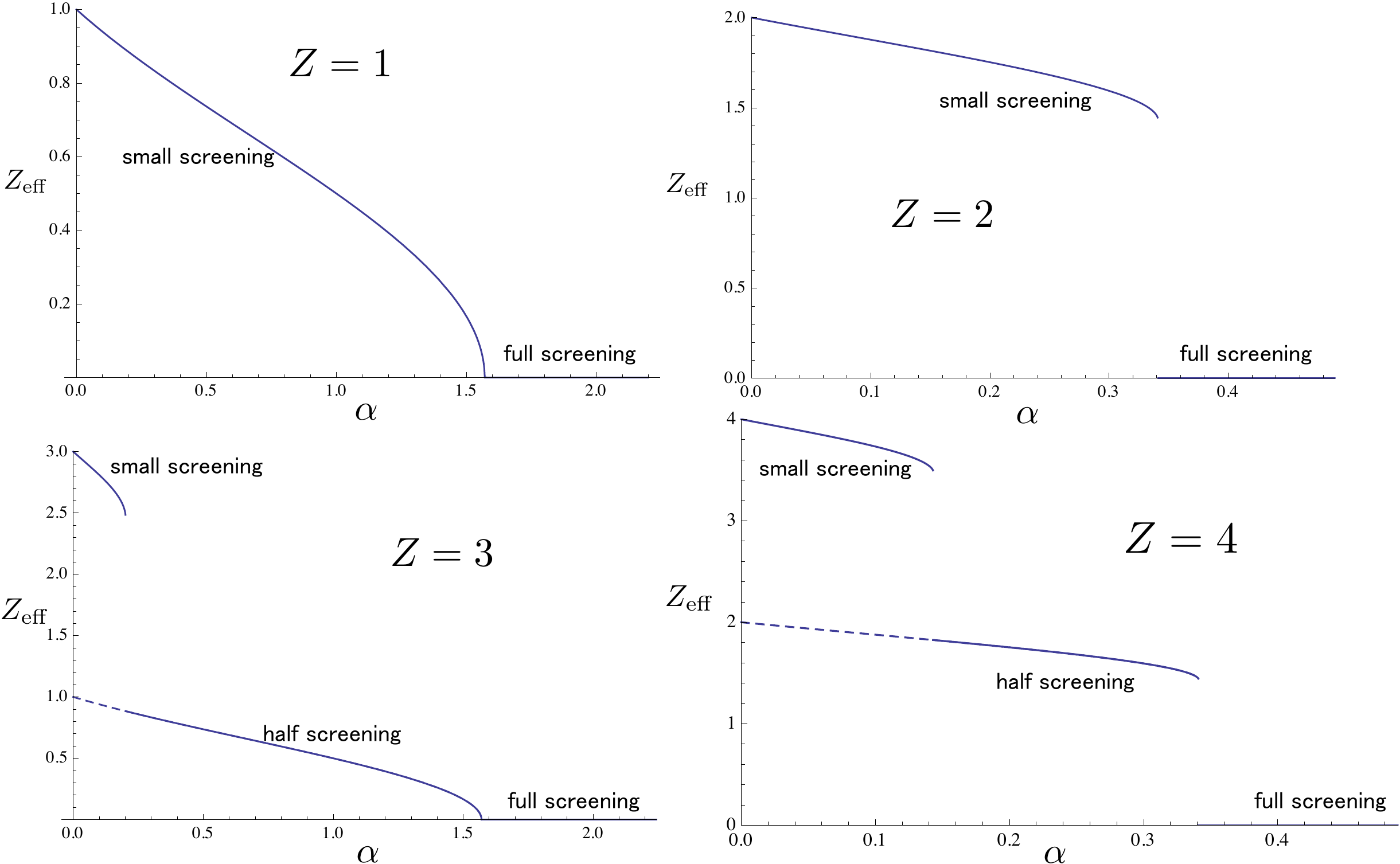}
		\caption{$Z_{\rm eff}$ for each $Z$. Dashed lines in $Z=3,4$ cases describe $Z_{\rm eff}$ in $Z=1,2$ cases, respectively.}
		\label{Zeff}
\end{figure}

\begin{figure}[htbp]
	\centering
	\includegraphics[width=12cm]{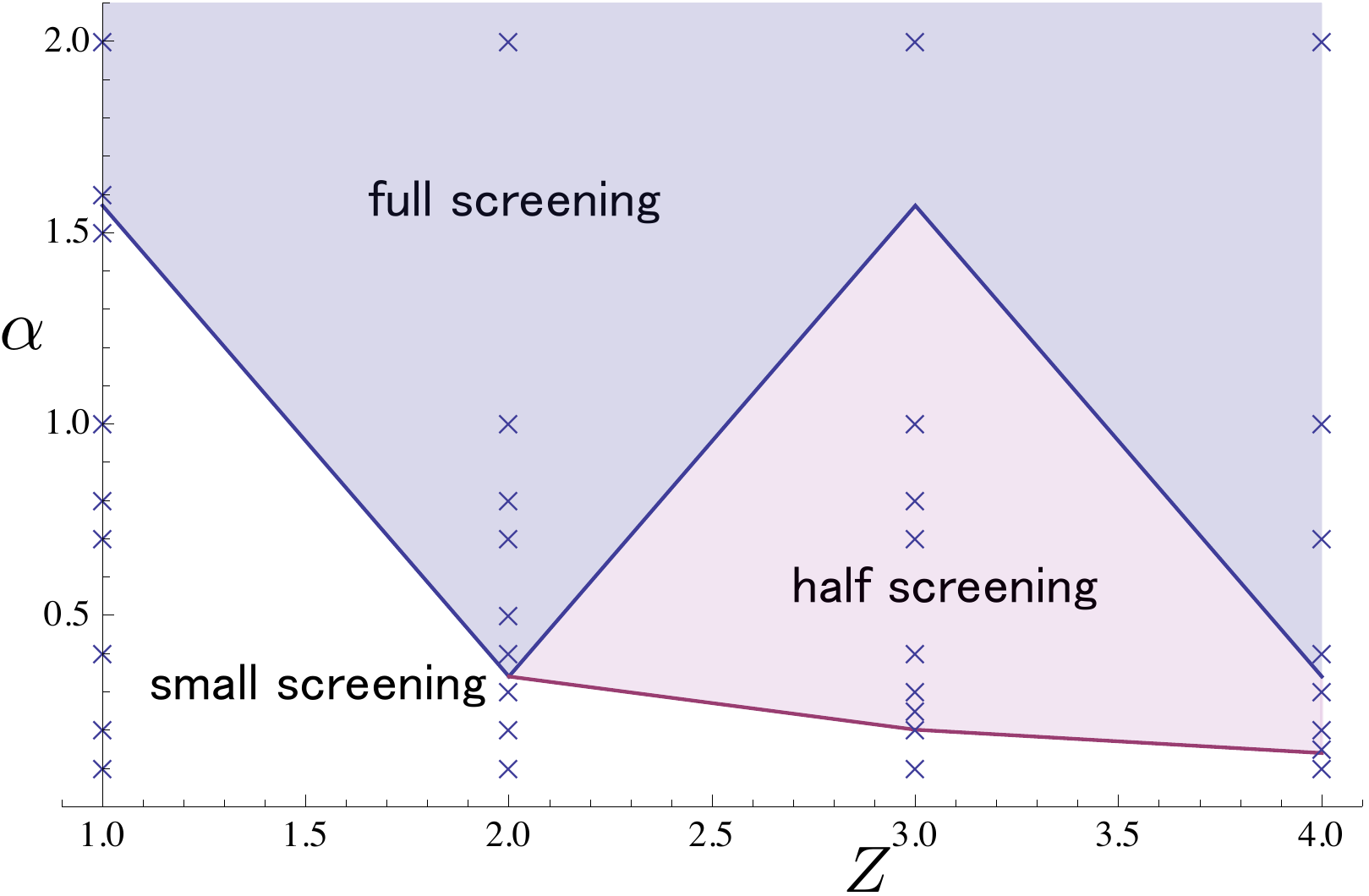}
	\caption{Phase diagram of screening. Crosses are the parameter points where we solved Eq. (\ref{ELEPhi+})}
	\label{phase}
\end{figure}

We looked for the solution for  various set of parameters of ($\alpha ,Z$), where the parameter set is given in Fig. \ref{Number} .
We found that only the solution with the smallest value of  $\phi_*$  can satisfy correct boundary condition at $r=0$ in all cases. 
 From this fact, we reach the conjecture that the magnitude of screening can be determined by the smallest intersection of $\sin (\sqrt{2\pi }\phi _*)$ and $-2\alpha ( \sqrt{2/\pi }\phi _*-Z) $. According to this conjecture, we get effective impurity charge seen from infinitely separated point,
\begin{align}
	Z_{\rm eff}=Z-\sqrt{\frac{2}{\pi }}\phi _*,
\end{align}
which is screened by induced charge Fig. \ref{Zeff}. Notice that when $\alpha \gtrsim 0.2$, the effective impurity charge $Z_{\rm eff}$ in any odd $Z$ case is the same one as in $Z=1$ case. Also when $\alpha \gtrsim 0.14$, $Z_{\rm eff}$ in any even $Z$ case is the same one as in $Z=2$ case. From this result, a phase diagram of screening is described as in Fig. \ref{phase}. In larger $Z$ case, more branches appear in small $\alpha $ regime. 
\subsubsection{Scaling law}
\begin{figure}[htbp]
		\centering
		\includegraphics[width=16cm]{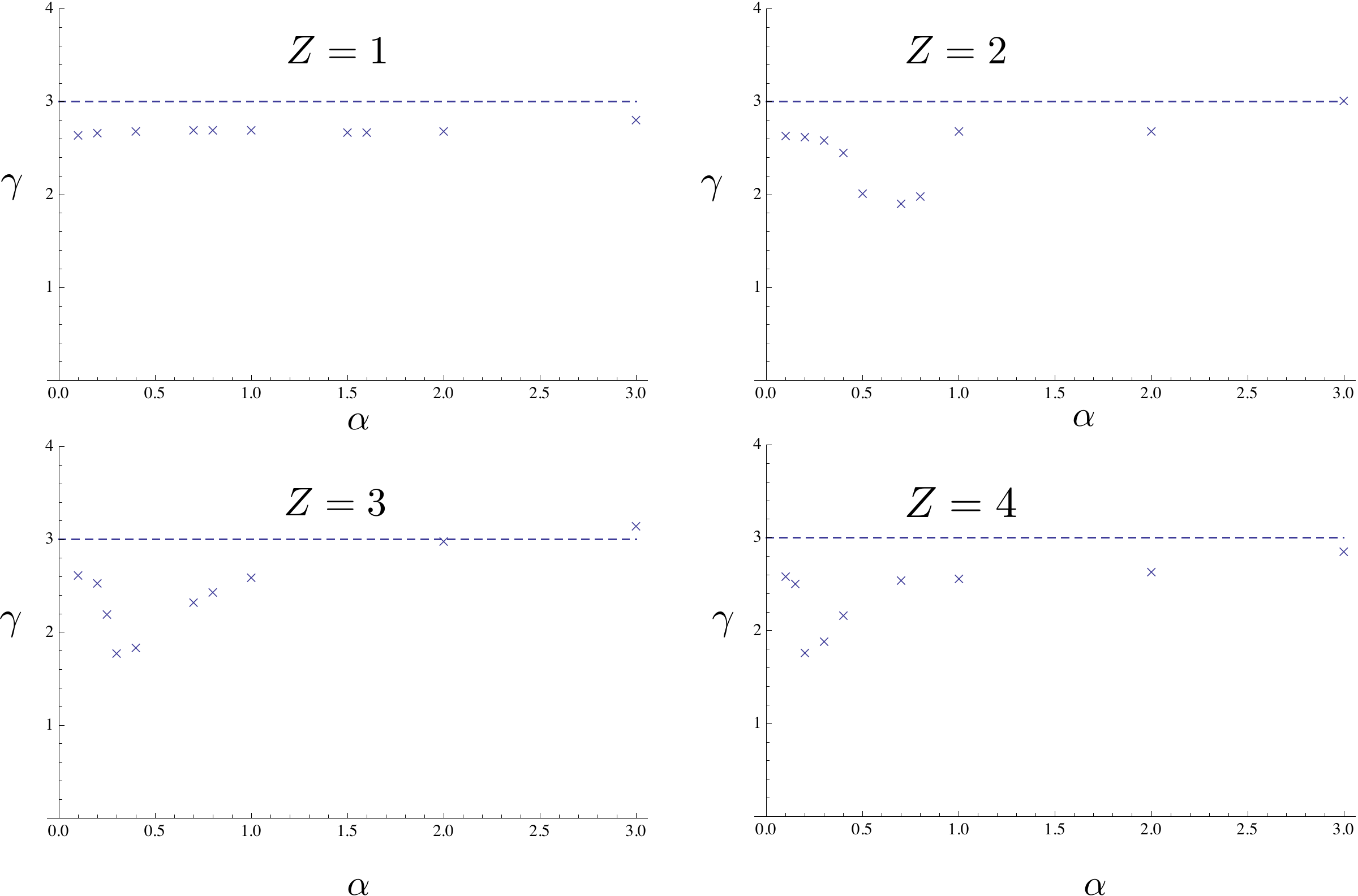}
		\caption{Crosses describe the scaling exponent in each $Z$ case. Dashed line describes one in perfect metal approximation.}
		\label{scaling}
\end{figure}

The induced 2D electron density can be obtained as
\begin{align}
	n(r)=\frac{\rho _e(r)}{2\pi r}.
\end{align}
If graphene sheet can be treated as perfect metal, the scaling law is calculated as in Ref. \cite{Fogler:2007}:
\begin{align}
	n(r)\propto r^{-3},
\end{align}
in the range of distances $1\ll r/R \ll 2\alpha ^2Z$. In our calculation, we fit the scaling law
\begin{align}
	n(r)\propto r^{-\gamma }
\end{align}
in the range of distances $1\ll r/R \ll 10$. The scaling exponent $\gamma $ depends on parameters $\alpha ,Z$ as shown in Fig.\ref{scaling}. In small screening regime, we get $\gamma \sim 2.7$, independently of $\alpha $. Near the value of $\alpha $ where magnitude of screening jumps, $\gamma $ drastically decreases. In larger $\alpha $ regime, $\gamma $ increases and becomes close to the value calculated in perfect metal approximation. 

%\input{4.3.Result.tex}
%\input{5.Result.tex}
%May 14 revised by AK
%June 11 revised by AK
%July 24 start revising AK, end revising AK
\section{Summary and Discussion}
\label{sec:discussion}
In this paper, we studied quantum field theory with the 2+1 dimensional massless fermion around an external Coulomb field. We reduced the theory to a two dimensional fermion theory, where the higher partial waves are neglected.  Bosonizing the theory, we have found the static solution of classical equation of motion for the boson field. The magnitude of screening is determined only by the asymptotic equation of motion. Which of these asymptotic solutions satisfies the boundary condition at $r=0$ is determined by dynamics. 

Through the study of several examples, we have concluded that the realized solution is always the smallest screening one. As a result, we have found patterns of screening depending on the coupling $\alpha $ and the impurity charge $Z$. The screening charge undergoes a drastic change as we change the value of $\alpha $ at some critical values. %There are jumps in $\alpha $ dependence of screening. 
We also obtained the phase diagram characterized by the patterns of screening. 

By solving the equation of motion in full spatial regime, we have obtained the spatial distribution of density of the induced electron. The radial profile of the two dimensional induced charge density can be fitted by negative power in $r$ which is the distance from the impurity. In weak coupling regime, scaling exponent $\gamma $ is independent of $\alpha $ and $Z$; $\gamma \approx 2.7$. Near the screening jumping point, $\gamma $ decreases. This means that the induced fermion is widely spread near the screening jumping point. And in larger $\alpha $ regime, $\gamma $ become close to the value of the perfect metal approximation; $\gamma \approx 3$.

The validity of the approximation to neglect higher partial wave can be discussed somewhat in semi classical theory mentioned in section 2. According to the semi classical theory, only $Z\alpha >j$ wave can form quasi-bound states. So, the fermion mode whose angular momentum $j$ is higher than $Z\alpha $ is irrelevant to anomalous behavior of the electron in strong Coulomb potential. When $Z\alpha >3/2$, the next to lowest partial wave $j=3/2$ should be relevant to this problem. Therefore our approximation should be valid only when $Z\alpha <3/2$. 

To compare our analysis with the result of one particle theory or the experiment, many things remain to be done. Validity of classical treatment for boson theory should be confirmed quantitatively. In Ref. \cite{Hirata:1989px}, the bosonized atomic collapse problem in 3+1 dimensions is treated within small fluctuation approximation. They show the existence of meta stable states in supercritical phase. In the same way it may be possible to show the existence of the meta stable states in our 2+1 dimensional massless fermion case.

The contribution of higher momentum partial wave should be evaluated for understanding larger $\alpha ,Z$ case. Furthermore to understand the behavior in the regime closer to the impurity, the effect of graphene lattice should be considered. For that purpose, the simulation by lattice gauge theory is important.

\acknowledgments
The authors would like to thank Masaki Hirotsu for discussions. 
This work was supported by  the Grant-in-Aid of the Japanese Ministry of Education (No. 26400248).

%\paragraph{Note added.} This is also a good position for notes added
%after the paper has been written.


%\begin{thebibliography}{99}

%\bibitem{a}
%Author, \emph{Title}, \emph{J. Abbrev.} {\bf vol} (year) pg.

%\bibitem{b}
%Author, \emph{Title},
%arxiv:1234.5678.

%\bibitem{c}
%Author, \emph{Title},
%Publisher (year).

% Please avoid comments such as "For a review'', "For some examples",
% "and references therein" or move them in the text. In general,
% please leave only references in the bibliography and move all
% accessory text in footnotes.

% Also, please have only one work for each \bibitem.

%\end{the bibliography}

\providecommand{\href}[2]{#2}\begingroup\raggedright\endgroup

\end{document}